\documentclass[12pt]{article}
\pdfoutput=1
\usepackage{latexsym, graphicx,cite}
\usepackage{amsmath}
\usepackage{amssymb}

\allowdisplaybreaks

\usepackage[top = 1. in,bottom = 0.93 in, left = 1 in, right=1 in]{geometry}
\usepackage[colorlinks=true, linkcolor=blue, bookmarks=true]{hyperref}
\newcommand{\arXiv}[1]{\href{http://www.arXiv.org/abs/#1}{arXiv:#1}}
\newcommand{\doi}[2]{\href{http://dx.doi.org/#2}{#1}}

\makeatletter
\renewcommand\section{\@startsection {section}{1}{\z@}%
                  {-3.5ex \@plus -1ex \@minus -.2ex}
                  {2.3ex \@plus.2ex}%
                  {\normalfont\large\bfseries}}
\renewcommand\subsection{\@startsection{subsection}{2}{\z@}%
                   {-3.25ex\@plus -1ex \@minus -.2ex}%
                   {1.5ex \@plus .2ex}%
                   {\normalfont\bfseries}}
\makeatother

\numberwithin{equation}{section}

\newcommand{\beq}{\begin{equation}}
\newcommand{\eeq}{\end{equation}}
\newcommand{\ber}{\begin{array}}
\newcommand{\eer}{\end{array}}
\newcommand{\del}{\partial}

\newcommand{\de}{\delta}

\newcommand{\bea}{\begin{eqnarray}}
\newcommand{\eea}{\end{eqnarray}}
\newcommand{\ER}{Erd\H{o}s-R\'enyi }

\newcommand{\nn}{\nonumber}

\begin{document}
\begin{titlepage}
\begin{flushright}
\phantom{arXiv:yymm.nnnn}
\end{flushright}
\vspace{5mm}
\begin{center}
{\LARGE\bf First return times on sparse random graphs}\\
\vskip 17mm
{\large Oleg Evnin$^{a,b}$ and Weerawit Horinouchi$^a$}
\vskip 7mm
{\em $^a$ High Energy Physics Research Unit, Department of Physics, Faculty of Science, Chulalongkorn University,
Bangkok, Thailand}
\vskip 3mm
{\em $^b$ Theoretische Natuurkunde, Vrije Universiteit Brussel and\\
International Solvay Institutes, Brussels, Belgium}
\vskip 7mm
{\small\noindent {\tt oleg.evnin@gmail.com, wee.hori@gmail.com}}
\vskip 20mm
\end{center}
\begin{center}
{\bf ABSTRACT}\vspace{3mm}
\end{center}

We consider random walks in the form of nearest-neighbor hopping on \ER random graphs of finite fixed mean degree $c$ as the number of vertices $N$ tends to infinity. In this regime, using statistical field theory methods, we develop an analytic theory of the first return time probability distribution. The problem turns out closely related to finding the spectrum of the normalized graph Laplacian that controls the continuum time version of the nearest-neighbor-hopping random walk. In the infinite graph limit, where loops are highly improbable, the returns operate in a manner qualitatively similar to $c$-regular trees, and the expressions for probabilities resemble those on random $c$-regular graphs. Because the vertex degrees are not exactly constant, however, the way $c$ enters the formulas differs from the dependence on the graph degree of first return probabilities on random regular graphs.

\vfill

\end{titlepage}

\section{Introduction}

Random walks on graphs are of relevance for phenomena as different as the flow of Internet traffic, the rise of popularity of social trends, and the spread of epidemics. The evolution of every random walk is marked by signature events: the first passage time (the number of steps it takes to reach a given node), the first return time (the number of steps it takes to revisit the starting position for the first time), the cover time (the number of steps it takes to visit every node).
The probability distributions controlling these special timepoints on a random network capture the key patterns in the random walk lifecycle.
A systematic exposition of related subjects can be found in \cite{rev1,rev2}.

In a recent series of works by Tishby, Biham and Katzav \cite{TBK1,TBK2,TBK3, TBK4}, these questions were addressed for random regular graphs. While in many contexts,
random regular graphs are far from being the simplest random graph ensemble (and indeed, even counting the total number of regular graph configurations is far from obvious \cite{Kawamoto}), this setting is very comfortable for analytic work on random walks. The reason is that the branching at each node is rigidly constrained by the specified vertex degree, while a typical large random regular graph is tree-like and lacks short loops. This leads to a simple picture, for example,
for the first return trajectories \cite{TBK2}. In the limit of an infinite graph, all such trajectories follow the `retroceding' scenario: the path must traverse each link ever visited an equal number of times in both directions. As a result, an elementary combinatorial estimate is possible for the probability of such `retroceding' returns on infinite random regular graphs, and even corrections to this picture for finite graphs can be understood to a significant extent \cite{TBK2}.

Our purpose here is to develop a similar analytic theory for (sparse) \ER random graphs, focusing on the distribution of first return probabilities.
(Other related quantities on \ER graphs have been considered in the past, for example, first passage times in \cite{frstpass} and hitting times in \cite{hitting}.) One usually thinks of \ER graphs as being the simplest possible random graph ensemble. All edges connecting $N$ vertices are filled independently with probabilities $c/N$, so that the mean vertex degree is $c$, and we refer to such graphs as `sparse' when $c\ll N$.
Despite this apparent simplicity, \ER graphs present a much more challenging setting for random walks than random regular graphs. On top of the randomness in where the walker goes at each next step, one encounters randomness in vertex degrees, which makes it impossible to directly apply combinatorial counting as in the considerations of \cite{TBK2}. The strategy of considering walks on tree-like graphs with a given degree distribution has been pursued in \cite{MK}. The techniques we shall resort to here often go under the name of statistical field theory methods \cite{statFT, statFTneu, largeN}, and they have been historically applied to a wide range of problems in random matrices and random graphs, see \cite{BR,RB,FM,MF,euclRM,SC,PN,kuehn,spectrum, AC,TO,resdist,RP2022,laplace} for a sampler of literature. These methods work best in the $N\to\infty$ limit, and for that reason we shall focus on infinite graphs.
Analysis of $1/N$ corrections is possible in some cases using this kind of techniques, as in \cite{spectrum}, for example, but it requires considerable effort.
More specifically, we shall be relying on the `supersymmetric' formulation \cite{efetov,wegner} for our statistical field theory that introduces auxiliary integrals
over anticommuting variables. This approach is very effective for handling random graph averages, and one can see a range of applications in \cite{FM,MF,TO,resdist,laplace}.

The random walks we consider proceed in discrete time steps, whereby the walker hops at each time step to one of the nearest neighbors of the current vertex with equal probability. The continuum analog of such random walks, where there is a fixed probability rate per unit time to leave the current vertex and move with equal probability to one of its neighbors, is known to be governed by the {\it normalized graph Laplacian}. It is not surprising then that our considerations will, at one point, approach very closely the solution of the spectral problem for the normalized Laplacian of \ER graphs developed in \cite{laplace} by methods very close to what we are planning to use here.

Our strategy can be described as follows: We shall first review, in section~\ref{seqgen}, the generating function for the first return probabilities, following the considerations of \cite{sites}. In section~\ref{seqaux}, we shall develop an auxiliary field representation for the generating function, introducing extra integrals in such a way that averaging over graphs becomes straightforward. In section~\ref{seqsddl}, we shall discuss the $N\to\infty$ saddle point structure in our auxiliary field representation, and direct contact will be made with the considerations of \cite{laplace}. In sections~\ref{seqest} and \ref{seqprobs}, we shall use the saddle point structure to develop accurate estimates for the first return probabilities (technically, in the asymptotic regime $1\ll c\ll N$). We shall thereafter conclude with a brief discussion and outlook in section~\ref{concl}. 

\section{The generating function of first return probabilities}\label{seqgen}

As customary in random graph theory, we shall encode the information about vertex connectivities by introducing the (symmetric) adjacency matrix $\mathbf{A}$, whose components $A_{ij}$ equal 1 if there is an edge between vertices $i$ and $j$, and 0 otherwise (while the diagonal components $A_{ii}$ are identically zero).
We furthermore introduce a diagonal matrix $\mathbf{D}$ encoding the vertex degrees whose only nonvanishing entries are 
\beq
D_{ii}\equiv \sum_j A_{ij}. 
\eeq
In the \ER random graph context, $A_{ij}$ will become independent identically distributed random variables for $i<j$.

Now consider a random walker that, at each time steps, jumps with equal probability to one of the nearest neighbors of its current vertex. If the current vertex is $i$, the probability to jump to vertex $j$ is evidently $A_{ji}/D_{ii}$. If, at time point $n$, the probability to be at vertex $i$ is $p_i(n)$, then the probability to be at vertex $i$ at time point $n+1$ is
\beq
p_i(n+1)= \sum_j \frac{A_{ij}}{D_{jj}} p_j(n).
\eeq
Then, if we start at vertex $j$ at $n=0$, the probability to be at vertex $j$ again at time point $n$ is given by
\beq\label{Gjjn}
G_{jj}(n)=[\mathbf{AD^{-1}}]^n_{jj},
\eeq
where the superscript $n$ denotes the ordinary matrix power, while $jj$ refers to the $jj$ component of that matrix. It will be convenient to consider the generating function for the probabilities (\ref{Gjjn}), given by
\beq\label{Gadj}
G_{jj}(z)\equiv\sum_{n=0}^\infty G_{jj}(n)z^n=[\mathbf{I}-z\mathbf{AD^{-1}}]^{-1}_{jj}.
\eeq

There is an elegant way to convert this generating function for probabilities to return to vertex $j$ to the generating function for probabilities of {\it first} returns to vertex $j$, which we borrow from \cite{sites}. (See \cite{rev2} for a textbook exposition of this topic.) We shall denote the probability of the first return to site $j$ after $n$ steps $F_{jj}(n)$. If the walker starts at vertex $j$ at moment zero and returns to this vertex at a later moment $n$, this can happen via one of the following mutually exclusive scenarios: either it is the zeroth return (being at the starting position), which is evidently only possible at $n=0$, or it is the first return after $n$ steps, or it is the second return consisting of a first return after $n_1$ steps and another first return after $n-n_1$ steps, and so on. Since all of these scenarios are mutually exclusive, the probability to return to vertex $j$ after $n$ steps is expressed as a sum of probabilities of all of these scenarios. In other words,
\begin{equation}\label{returndecomp}
    G_{jj}(n)=\delta_{n,0}+F_{jj}(n)+\hspace{-2mm}\sum_{0\le n_{1}\le n}\hspace{-2mm}F_{jj}(n)F_{jj}(n-n_{1})+\hspace{-5mm}\sum_{0\le n_{1}+n_2\le n}\hspace{-5mm}F_{jj}(n_1)F_{jj}(n_2)F_{jj}(n-n_{1}-n_2)+\cdots
\end{equation}
Introducing the generating function
\beq
F_{jj}(z)\equiv \sum_{n=0}^\infty z^n F_{jj}(n),
\eeq
we can equivalently rewrite (\ref{returndecomp}) in a much more compact way by multiplying it with $z^n$ and summing over $n$, which yields
\beq
    {G}_{jj}(z)=1+{F}_{jj}^{}(z)+{F}_{jj}^{2}(z)+\cdots=\frac{1}{1-{F}_{jj}(z)}. \label{G1}
\eeq
We then simply invert this relation to express $F_{jj}(z)$, which is the main target of our analytic efforts, through $G_{jj}(z)$, for which an explicit expression (\ref{Gadj}) in terms of the adjacency matrix is available:
\begin{equation}\label{Fjjadj}
    F_{jj}(z)=1-\frac{1}{G_{jj}(z)}=1-\frac{1}{[\mathbf{I}-z\mathbf{AD^{-1}}]^{-1}_{jj}}.
\end{equation}

\section{An auxiliary field representation}\label{seqaux}

Summed up in one sentence, the mathematical content of this work is in averaging (\ref{Fjjadj}) over adjacency matrices corresponding to \ER graphs, or more specifically, treating $A_{ij}$ with $i<j$ as independent random variables equal 1 with probability $c/N$ and 0 with probability $1-c/N$. Since the ensemble is invariant under relabelling the graph vertices, it is evident that
\beq
\langle F_{jj}(z)\rangle=\langle F_{11}(z)\rangle
\eeq
for any $j$, and hence we shall formulate the rest of our computation in terms of $F_{11}(z)$. Note that the average we perform does not make any prescriptions about whether vertex 1 belongs to the giant component of the graph or to small components, and includes all these cases indiscriminately.

Averaging (\ref{Fjjadj}) is evidently something that is easier said than done, since the dependence on the adjacency matrix is very complicated.
The key idea of auxiliary field methods is to re-express (\ref{Fjjadj}) using extra integrals to make the dependence on $A_{ij}$ simplify.
If one pushes the expressions to a form where they factorize over $A_{ij}$, the averaging over graphs can be straightforwardly performed.
In practice, this factorization happens when, after a number of auxiliary integrals are introduced, the integrand become an exponential of something linear in $A_{ij}$. Such methods are widely applied in statistical physics and in random matrix theory, but in our case, the intricate expression (\ref{Fjjadj}) prompts
a relatively sophisticated auxiliary field treatment.

First of all, to facilitate subsequent contacts with the spectral problem of the normalized graph Laplacian, we shall introduce
\beq
w\equiv 1/z
\eeq
and rewrite (\ref{Fjjadj}) as
\begin{equation}\label{Fjjw}
    F_{11}(z)=1-\frac{z}{[w\mathbf{I}-\mathbf{AD^{-1}}]^{-1}_{11}}=1+\frac{z}{D_{11}[\mathbf{A}-w\mathbf{D}]^{-1}_{11}}.
\end{equation}
Note the appearance of the resolvent-like combination $[\mathbf{A}-w\mathbf{D}]^{-1}$.
We then introduce Schwinger-like parametrizations ($a^{-1}=\int_0^\infty dt e^{-at}$) for the inverses in the second term: 
\begin{equation}
    F_{11}(z)=1-iz\int_{0}^{\infty} \!\!ds \int_{0}^{\infty} \!\!dt\; e^{is\,{[\mathbf{A}-w\mathbf{D}]^{-1}_{11}}}e^{-tD_{11}}. 
\label{FSch}
\end{equation}
The insertion of the imaginary $i$ factors is motivated by making contacts with the results of \cite{laplace} at a later point
in our derivations. 

For the purposes of averaging (\ref{FSch}), it is convenient to introduce
\beq\label{curlF}
\mathcal{F}(s,t,w)\equiv \langle e^{is\,{[\mathbf{A}-w\mathbf{D}]^{-1}_{11}}}e^{-tD_{11}}\rangle.
\eeq
We shall formulate all of our computations in terms of this quantity, while the average of (\ref{FSch}) can be recovered at the end of the day as
\beq\label{FcurlF}
    \langle F_{11}(z)\rangle=1-iz\int_{0}^{\infty} \!\!ds  \int_{0}^{\infty} \!\! dt \;  \mathcal{F}(s,t,1/z).
\eeq

The expression (\ref{curlF}) is getting close to something one could compute, but direct averaging over $A_{ij}$ is still impossible because of the appearance
of the resolvent in the exponent. One can introduce, however, further Gaussian integrals to remedy for that in a manner closely parallel, for example,
to the considerations of \cite{laplace}, where one had to deal with averaging the resolvent itself, rather than its exponential.

As a starter, observe that, for any matrix $\mathbf{M}$, introducing a vector $\pmb{\phi}\equiv(\phi_1,\phi_2,\ldots, \phi_N)$,  we can write
\begin{equation}\label{Gauss}
 e^{is{M^{-1}_{11}}}\propto \sqrt{\det\mathbf{M}}\int d\pmb{\phi} \; \exp\left(i\sum_{kl}\phi_{k}M_{kl}\phi_{l}\right) e^{2i\sqrt{s}\phi_{1}} .
\end{equation}
If one takes, $\mathbf{M}=\mathbf{A}-w\mathbf{D}$ and uses this representation in (\ref{curlF}), the structure is quite promising in that one gets an exponential of something linear in $A_{ij}$, so that a factorization over the components of $\mathbf{A}$ occurs. The problem is in the determinant prefactor in (\ref{Gauss}), which does not factorize. Note that we have used the proportionality sign in (\ref{Gauss}) to indicate the omission of purely numerical factors independent of $s$. We shall employ a similar strategy in what follows since such purely numerical factors would be cumbersome to keep. The overall numerical normalization will be fixed at the end of the computation by an independent argument that we shall specify below.

A powerful way to deal with unwanted determinant factors as in (\ref{Gauss}) is to introduce a few extra Gaussian integrals, including those over anticommuting (Grassmannian) variables, so that the determinant factors cancel among these integrals, leaving behind only exponentials linear in $A_{ij}$. This is known as the supersymmetry-based approach to disorder averaging \cite{MF,FM,resdist,laplace,efetov, wegner}.
We introduce two $N$-vectors $\pmb{\xi}$ and $\pmb{\eta}$ with anticommuting (Grassmannian) components $\xi_i$ and $\eta_i$ that satisfy the relations
\beq\label{Berezin}
\xi_i\xi_j=-\xi_j\xi_i,\qquad \xi_i^2=0,\qquad \int d\xi_i \xi_j=\de_{ij},\qquad \int d\xi_i=0,
\eeq
and similar relations for $\eta_i$. Furthermore, all components of $\pmb{\xi}$ also anticommute with the components of $\pmb{\eta}$. The integration rules written above, in particular, ensure that the following Gaussian integration formula are valid for any $N\times N$ matrix $\mathbf{M}$:
\beq\label{Berezindet}
\det \mathbf{M}\propto \int d\pmb{\xi}\, d\pmb{\eta}\, e^{i\sum_{kl}\xi_k{M}_{kl}\eta_l}.
\eeq
If we finally incorporate an `empty' Gaussian integral over an $N$-component vector $\pmb{\chi}$ made of ordinary commuting numbers,
\beq\label{Gausschi}
1\propto \sqrt{\det\mathbf{M}}\int d\pmb{\chi} \, e^{i\sum_{kl}\chi_k M_{kl} \chi_l},
\eeq
the determinant factors cancel each other, leaving behind a much more useful version of (\ref{Gauss}) in the form
\begin{equation}\label{SGauss}
 e^{is{M^{-1}_{11}}}\propto \int d\pmb{\phi}  d\pmb{\chi}   d\pmb{\xi}   d\pmb{\eta}  \; e^{i\sum_{kl}M_{kl}(\phi_{k}\phi_{l}+\chi_{k}\chi_{l}+\xi_{k}\eta_{l})} e^{2i\sqrt{s}\phi_{1}}.
\end{equation}
(The integrals over $\pmb{\xi}$ and $\pmb{\eta}$ contribute $\det \mathbf{M}$ according to (\ref{Berezindet}), while the integrals over $\pmb{\phi}$
and $\pmb{\chi}$ contribute $1/\sqrt{\det \mathbf{M}}$ each, leaving no powers of $\det \mathbf{M}$ left.)
It is convenient to introduce the following `superspace' notation: we join $\phi_k$, $\chi_k$, $\xi_k$ and $\eta_k$ into a `supervector' 
\beq\label{Psicomp}
\Psi_k=(\phi_k, \chi_k, \xi_k, \eta_k)^T,
\eeq
whose conjugate $\Psi_k^\dagger$  is defined using the `supermetric' $S$ as
\beq
\Psi_k^\dagger\equiv \Psi_k^T\,S,\qquad S\equiv\left(\begin{matrix}\hspace{2mm}1\hspace{2mm}&0&0&0\\0&1&0&0\\0&0&0& -1/2\\0&0&1/2&0\end{matrix}\right).
\eeq
Then,
\begin{equation}\label{Gausssuperv}
 e^{is{M^{-1}_{11}}}\propto \int d\pmb{\Psi}\, e^{i\sum_{kl}M_{kl} \Psi^\dagger_k \Psi_l} e^{2i\sqrt{s}\phi_{1}} .
\end{equation}
Substituting  $\mathbf{M}=\mathbf{A}-w\mathbf{D}$ and keeping in mind that
\beq
\sum_{kl} (A_{kl}-w D_{kl}) x_k x_l = \sum_{kl} A_{kl} x_k x_l - w\sum_{kl} A_{kl}   x_k^2 =\sum_{k<l} A_{kl} (2x_kx_l-wx_k^2-wx_l^2),
\eeq
we obtain\footnote{We are being somewhat careless in inverting $\mathbf{A}-w\mathbf{D}$, since any degree 0 vertex contributes a null eigenvalue of this matrix that would, at the face value, lead to divergences. However, the probability for a vertex to have degree 0 is $\sim e^{-c}$, hugely suppressed at large $c$. Since we will be focusing on large $c$ and $1/c$ corrections (which is the only currently viable approach for doing analytics), the issue will not be relevant, and no divergences will be seen. This is very similar to the related computations in \cite{laplace}.\label{footnoteec}}
\beq\label{SGausgrph}
 e^{is\,{[\mathbf{A}-w\mathbf{D}]^{-1}_{11}}}\propto  \int d\pmb{\Psi}\, e^{i\sum_{k<l}A_{kl} (2\Psi^\dagger_k \Psi_l-w\Psi^\dagger_k \Psi_k-w\Psi^\dagger_l \Psi_l)} e^{2i\sqrt{s}\phi_{1}}.
\eeq
This expression is both fully factorized over the entries of $\mathbf{A}$ and depends explicitly only on the independent entries $A_{kl}$ with $k<l$. It is thus fully prepared for averaging over graphs.

When we substitute (\ref{SGausgrph}) into (\ref{curlF}), and take into account that $D_{11}\equiv A_{12}+A_{13}+\cdots+A_{1N}$, we get
\beq\label{SGausgrph1}
\mathcal{F}(s,t,w)\propto  \int d\pmb{\Psi}\,  e^{2i\sqrt{s}\phi_{1}}\langle \prod_{j\ge 2} e^{-tA_{1j}} \prod_{k<l}e^{i A_{kl} (2\Psi^\dagger_k \Psi_l-w\Psi^\dagger_k \Psi_k-w\Psi^\dagger_l \Psi_l)}\rangle.
\eeq
As $A_{kl}$ with $k<l$ are independent and equal 1 with probability $c/N$ and 0 with probability $1-c/N$, the average is evaluated straightforwardly to yield
\begin{align}\label{SGausav}
\mathcal{F}(s,t,w)\propto  \int d\pmb{\Psi}\,  e^{2i\sqrt{s}\phi_{1}} &\prod_{j\ge 2}\left\{1+\frac{c\,}{N}\left[e^{-t}e^{i  (2\Psi^\dagger_1 \Psi_j-w\Psi^\dagger_1 \Psi_1-w\Psi^\dagger_j \Psi_j)}-1\right]\right\}\\
&\times\prod_{2\le k<l}\left\{1+\frac{c}{N}\left[e^{i  (2\Psi^\dagger_k \Psi_l-w\Psi^\dagger_k \Psi_k-w\Psi^\dagger_l \Psi_l)}-1\right]\right\}.\nn
\end{align}
Since $c/N$ becomes very small at $N\to \infty$, using $1+x/N\approx \exp(x/N)$, this may be rewritten as
\begin{align}
\mathcal{F}(s,t,w)\propto \exp[-c(1-e^{-t})]  \int d\pmb{\Psi}\,& e^{2i\sqrt{s}\phi_{1}}
\exp\Big[-\frac{e^{-t}}N\sum_{j\ge 2}C(\Psi_1,\Psi_j)-\frac{1}{2N}\sum_{k,l\ge 2}C(\Psi_k,\Psi_l)\Big]\nn\\
&\times\exp\Big[\frac{1}{2N}\sum_{k}C(\Psi_k,\Psi_k)\Big],
\label{SGausavC}
\end{align}
with
\beq\label{Cdef}
 C(\Psi,\Psi')\equiv c\left(1-e^{i  (2\Psi^\dagger \Psi'-w\Psi^\dagger \Psi-w\Psi'^\dagger \Psi')}\right).
\eeq
To obtain (\ref{SGausavC}) from (\ref{SGausav}), we represented the last $-1$ in the first line of (\ref{SGausav}) as $-1=-e^{-t}-(1-e^{-t})$, and took the contribution due to $(1-e^{-t})$ out of the $j$-product, combining them into the overall prefactor $\exp[-c(N-1)(1-e^{-t})/N] \approx \exp[-c(1-e^{-t})] $. In the second line of (\ref{SGausavC}), we collected the diagonal contributions that arise from converting the sum over $k<l$ in (\ref{SGausav}) to a sum without this condition
in (\ref{SGausavC}). These contributions will play no role in the final answer and will drop out from our considerations shortly.

We need to make a remark on fixing the overall normalization in (\ref{SGausavC}). Determining numerical prefactors using statistical field theory techniques is a tedious task that requires keeping track of purely numerical factors in various integration measures. Since the uncertainty is in a single numerical constant, it is wise to compute (\ref{SGausavC}) up to normalization, which is what we shall do, and fix the normalization in the end using some known exact property of return probabilities. The easiest approach of this sort is to observe that $F_{jj}(z=0)=0$ according to (\ref{Fjjadj}). This relation holds for any given graph, expressing the tautology that the first return cannot occur after zero steps, and therefore it certainly holds after averaging over graphs. Thus, when we compute $\mathcal F$ from (\ref{SGausavC}) and substitute it into (\ref{FcurlF}), the normalization of $\mathcal F$ should be chosen such that  $\langle F_{11}(z=0)\rangle=0$. This will fix the last remaining ambiguity in our evaluation.

\section{The saddle point structure}\label{seqsddl}

We now turn to evaluating (\ref{SGausavC}) in the $N\to\infty$ limit. A practical problem with that expression is that the $C$-term in the exponent couples different $\Psi_k$. (If, by contrast, we had, instead of the $C$-term, a sum of terms each of which depended on only one $\Psi_k$, the integrand would factorize and the integrations could be immediately performed.)

Fyodorov and Mirlin proposed in \cite{FM,MF} an elegant trick to deal with integrals of the type of (\ref{SGausavC}). The idea is to introduce further auxiliary integrals in a manner that makes the integrand factorize over $\Psi_k$. While the price for that is that functional integrals enter the stage, once this has happened, a saddle point structure at large $N$ immediately emerges, leading to an integral saddle point equation whose solution encodes the $N\to\infty$ estimate of (\ref{SGausavC}).

In more conventional applications of the Fyodorov-Mirlin method \cite{FM,MF,laplace}, one makes use of the following functional Gaussian integral
\beq\label{MFtransfrm}
\int \mathcal{D}g \exp\left[-\frac{N}2\int d\Psi\,d\Psi'\, g(\Psi) \,C^{-1}(\Psi,\Psi')\,g(\Psi')+i\sum_k g(\Psi_k)\right]=e^{-\frac1{2N}\sum_{kl}C( \Psi_k, \Psi_l)}.
\eeq
Here $g(\Psi)$ is a number-valued function of its supervector argument, and
 $C^{-1}$ is the inverse of $C$ defined as
\beq
\int d\Psi\, C(\Psi_1,\Psi)\,C^{-1}(\Psi,\Psi_2)=\de(\Psi_1-\Psi_2).
\eeq
Note that the $\de$-function of a supervector $\Psi=(\phi,\chi,\xi,\eta)^T$ is defined as
\beq
\de(\Psi)=-\xi\,\eta\,\de(\phi)\de(\chi),
\eeq
so that
\beq
\int d\Psi' F(\Psi') \,\de(\Psi-\Psi')=F(\Psi)
\eeq
for any $F$, cf.\ the integration rules (\ref{Berezin}). 

To give the reader some intuition about the Gaussian functional integral formula (\ref{MFtransfrm}), we compare it with the simplest Gaussian functional integral over a function $f(x)$ of an ordinary variable $x$ with a linear source $J(x)$:
\begin{align*}
&\int \mathcal{D}f \exp\left[-\frac12 \int dx\,dx'\, f(x)K(x,x')f(x')+i\int dx\, J(x)f(x)\right]\\
&\hspace{8cm}=\exp\left[-\frac12 \int dx\,dx'\, J(x)K^{-1}(x,x')J(x')\right].
\end{align*}
(This Gaussian integration underlies the standard textbook constructions of perturbative quantum field theory.)
To go from this formula to (\ref{MFtransfrm}), we need to substitute $J(x)=\sum_k \de(x-x_k)$, and then proceed with replacing ordinary variables $x$ with supervectors $\Psi$, the corresponding functions $f(x)$ with $g(\Psi)$, and $K$ by $C^{-1}$. The above functional integral over $f$ is furthermore a direct functional analog of the ordinary multivariate Gaussian integral with a linear source $z$,
\beq
\int [dy] \exp\left[-\frac12\sum_{kl} y_k K_{kl} y_l +i\sum_k y_k z_k\right]=\exp\left[-\frac12\sum_{kl} z_k K^{-1}_{kl} z_l \right],
\eeq
which can be immediately proved by rotating the coordinates to the eigenbasis of $K$. In all the above formulas, we have absorbed the determinant factors arising from Gaussian integrations (which do not depend, respectively, on $z$, $J$, $\Psi$) into the definition of the measure.

The structure in (\ref{MFtransfrm}) is evidently close to (\ref{SGausavC}), but there is one difference: where one of the arguments of $C$ in (\ref{SGausavC}) is $\Psi_1$, $C$ gets decorated with a prefactor of $e^{-t}$. Incorporating this feature simply amounts to inserting $e^{-t}$ in front of $g(\Psi_1)$ in (\ref{MFtransfrm}) to obtain
\begin{align}\label{MFmod}
\int \mathcal{D}g \exp\Big[-\frac{N}2\int d\Psi\,d\Psi'\, &g(\Psi) \,C^{-1}(\Psi,\Psi')\,g(\Psi')+ie^{-t}g(\Psi_1)+i\sum_{k\ge 2} g(\Psi_k)\Big]\\
&=\exp\Big[-\frac{e^{-t}}N\sum_{j\ge 2}C(\Psi_1,\Psi_j)-\frac{1}{2N}\sum_{k,l\ge 2}C(\Psi_k,\Psi_l)\Big],\nn
\end{align}
where we have dropped the diagonal contribution $e^{-2t}C(\Psi_1,\Psi_1)/N$, since it is just a single term explicitly suppressed by $1/N$, so that it cannot contribute at $N\to\infty$ (while, by contrast, a sum of, say, $N$ such terms can contribute).

When (\ref{MFmod}) is substituted into (\ref{SGausavC}), the ${\Psi_k}$-integrals will completely factorize, yielding $N-1$ identical factors, while the integral over $\Psi_1$ is somewhat modified.
The result can be recast in the following form:
\beq
\label{afterMF}
\begin{split}
&\mathcal{F}(s,t,w)\propto  \exp[-c(1-e^{-t})]\int \mathcal{D}g\,e^{-N\hspace{0.1mm}S[g]}\,
\frac {\int d\Psi \,e^{2i\sqrt{s}\phi}\,\exp\left[ie^{-t}g(\Psi)\right]}{\int d\Psi\,e^{ig(\Psi)}},\\
&S[g]\equiv\frac{1}2\int d\Psi\,d\Psi'\, g(\Psi) \, C^{-1}(\Psi,\Psi')\,g(\Psi')-\ln\left(\int d\Psi\,e^{ig(\Psi)+C(\Psi,\Psi)/2N}\right).
\end{split}
\eeq
The contribution $C(\Psi,\Psi)/2N$ arises from the diagonal terms in the second line of (\ref{SGausavC}). It is suppressed by an explicit power of $1/N$ and amounts, in the end, to an irrelevant $1/N$ correction to $S$ that will not affect the final result.

There is an evident `saddle point' structure in (\ref{afterMF}) due to the presence of the large factor $N$ in the exponent, and one expects that the integral is dominated at large $N$ by the stationary points of the functional $S[g]$ defined by $\de S/\de g(\Psi)=0$, which can be written out explicitly as
\beq\label{sddlpsi}
g(\Psi)=i\,\frac{\int d\Psi'\,C(\Psi,\Psi')\,e^{ig(\Psi')}}{\int d\Psi'\,e^{ig(\Psi')}}.
\eeq
(The term $C(\Psi,\Psi)/2N$ in (\ref{afterMF}) does not contribute to the saddle point equation, as it is suppressed by $1/N$.)
This equation is in fact identical to the one obtained in \cite{laplace} while solving the problem of eigenvalue density of the normalized Laplacian of the \ER graph. By contrast, the specific insertion multiplying the saddle point exponential in (\ref{afterMF}) is different from the considerations of \cite{laplace}.

Equations similar to (\ref{sddlpsi}) occur very commonly when supersymmetry-based methods are used for analyzing random matrix problems \cite{FM,MF,laplace}. They also show some intriguing resemblance to analogous equations emerging from the replica approach to similar problems \cite{BR,RB}, where instead of supervectors $\Psi$ one deals with multi-component vectors made from commuting variables.

A common strategy to deal with such equations is to observe the symmetry with respect to ``superrotations'' of $\Psi$ that preserve the inner product $\Psi_1^\dagger\Psi_2$. One can then look for solutions that are themselves symmetric (an alternative would have been families of saddle point configurations related by superrotations). In other works, we impose
\beq\label{gsymm}
g(\Psi)=g_*(\Psi^\dagger\Psi), \qquad \Psi^\dagger\Psi\equiv \phi^2+\chi^2+\xi\eta,
\eeq
and explore the consequences.
For functions of the form (\ref{gsymm}), the saddle point equation can be simplified. One introduces the polar coordinates in the $(\phi,\chi)$-plane,
\beq\label{polar}
\rho=\phi^2+\chi^2,\qquad \phi=\sqrt{\rho}\cos\alpha,\qquad \chi=\sqrt{\rho}\sin\alpha,\qquad d\phi\,d\chi=\frac12 d\rho\, d\alpha.
\eeq
and proceeds performing the integration over anticommuting ($\xi$,$\eta$) and angular ($\alpha$) variables, obtaining in the end
\begin{equation}\label{sddlrho}
    g_*(\rho)=-ic\left(e^{-i\rho w}-1-\rho\, e^{-i\rho w}\int_0^\infty d\rho' \; \frac{J_{1}(2\sqrt{\rho\rho'})}{\sqrt{\rho\rho'}}e^{ig_*(\rho')-i\rho'w}\right).
\end{equation}
Here,  $J_{1}(z)=\frac{1}{2\pi}\int_{0}^{2\pi} d\theta \,\cos\theta\,e^{iz\cos\theta}$ is the Bessel function. Equation (\ref{sddlrho}) is identical (up to the replacement $w\to z-1$) to the saddle point equation emerging from the problem of finding the spectral density of normalized graph Laplacians in \cite{laplace}, where further technical details of the derivation can be consulted. The way this solution is processed to yield the final answer is rather different for the setting of \cite{laplace} and for our present setting, and we shall deal with these matters below. Similar integral equations involving Bessel functions have been seen for a range of related spectral problems on sparse random graphs \cite{BR,RB,FM,MF,gel,Khetal,inteq}. The emergence of normalized Laplacian from related random walk problems can also be seen in a mathematically rigorous context of \cite{normlap}, where \ER graphs in the connected regime $c\ge\log N$ are considered.

Being an integral equation with an exponential nonlinearity, (\ref{sddlrho}) is of course daunting. We shall in practice resort to building its solutions asymptotically at large $c$. Establishing a more comprehensive way to handle such equations is a valuable objective for future developments. We note that rather similar equations have been handled numerically in \cite{gel}.

Once a solution (or an approximate solution) of (\ref{sddlrho}) has been obtained, one needs to derive the corresponding saddle point estimate of (\ref{afterMF}). As explained in \cite{laplace}, for supersymmetric $g$ of the form (\ref{gsymm}), the saddle point value of the action $S[g_*]$, the functional determinant at the saddle point, and the integral $\int d\Psi\,e^{ig(\Psi)}$ in the denominator all provide only irrelevant $w$-independent numerical constants (remember that we have a way to normalize $\mathcal{F}$ at the end of the computation, so purely numerical factors in all intermediate computations can be safely dropped). We thus obtain a saddle point estimate of the form
\beq\label{ins}
\mathcal{F}(s,t,w)\propto  \exp[-c(1-e^{-t})]\int d\Psi \,e^{2i\sqrt{s}\phi}\,\exp\left[ie^{-t}g_*(\Psi^\dagger\Psi)\right].
\eeq
This expression is substituted into (\ref{FcurlF}) to obtain
\beq\label{Fest}
    \langle F_{11}(z)\rangle=1+\frac{z\,\mathcal{N}}{\pi}\int_{0}^{\infty} ds  \int_{0}^{\infty} dt \exp[-c(1-e^{-t})]\int d\Psi \,e^{2i\sqrt{s}\phi}\,\exp\left[ie^{-t}g_*(\Psi^\dagger\Psi)\right],
\eeq
where $g_*$ depends on $w\equiv 1/z$ implicitly via the saddle point equation (\ref{sddlrho}), and $\mathcal{N}$ is a normalization factor\footnote{Inserting the factor of $1/\pi$ in (\ref{Fest}) is purely a matter of convenience since it could in any case be absorbed into the yet-unknown normalization $\mathcal{N}$. The origin of this factor, nonetheless, may be optionally understood from the denominator in the first line of (\ref{afterMF}), the fact that $g_*(0)=0$ as follows from (\ref{sddlrho}), and the integration identity $\int d\Psi F(\Psi^\dagger\Psi)=\pi F(0)$. The same integration identity assures that the supersymmetric term $C(\Psi,\Psi)/2N$ in (\ref{afterMF}) cannot contribute any nontrivial $(s,t,w)$-dependence in (\ref{ins}), given that the saddle point configuration of $g(\Psi)$ is supersymmetric.} derived as explained at the end of section \ref{seqaux}. To simplify this expression, recall that a supersymmetric $g_*(\Psi^\dagger\Psi)$ depends on the $\phi$ and $\chi$ components of $\Psi$ only through the rotationally invariant combination $\phi^2+\chi^2$. Then, as a warm-up, consider the integral $\int ds dx dy \exp(2i\sqrt{s} x) f(x^2+y^2)$. We can always introduce a dummy integral over the angle $\alpha$ that the integrand does not depend on, $\int d\alpha/2\pi$. Then, one can rotate $x$ and $y$ through the angle $\alpha$ to write
\begin{align}
& \int_0^\infty \hspace{-2mm}ds \int_{-\infty}^\infty \hspace{-3mm} dx \int_{-\infty}^\infty  \hspace{-3mm}dy\, e^{2i\sqrt{s} x} f(x^2+y^2)= \frac1{2\pi}\int_0^{2\pi} \hspace{-3mm} d\alpha\int_0^\infty \hspace{-3mm} ds \int_{-\infty}^\infty \hspace{-3mm} dx \int_{-\infty}^\infty \hspace{-3mm} dy\,  e^{2i\sqrt{s} x} f(x^2+y^2)\nn\\
&\hspace{2cm}= \frac1{2\pi}\int_0^{2\pi} \hspace{-3mm} d\alpha\int_0^\infty \hspace{-3mm} ds \int_{-\infty}^\infty \hspace{-3mm} dx \int_{-\infty}^\infty \hspace{-3mm} dy\,  e^{2i\sqrt{s} (x\cos\alpha+y\sin\alpha)} f(x^2+y^2)\nn\\
&\hspace{4cm}=\frac1{4\pi}\int_{-\infty}^\infty \hspace{-3mm}\,dk_x\,dk_y\,dx\, dy\,  e^{i\vec{k}\cdot\vec{x}} f(x^2+y^2)=\pi f(0),\label{FourFour}
\end{align}
where we have introduced $\vec{x}\equiv(x,y)$, $\vec{k}\equiv(k_x,k_y)$, $k_x\equiv 2\sqrt{s}\cos\alpha$, $k_y\equiv 2 \sqrt{s}\sin\alpha$. The integrals in the last line look exactly like a Fourier transform of $f(x^2+y^2)$, followed by an inverse Fourier transform, and so it returns, up to a constant factor, the value of $f(x^2+y^2)$ at $x=y=0$. In application of (\ref{FourFour}) to (\ref{Fest}), the value of $\Psi^\dagger\Psi$ at $\phi=\chi=0$ is $\xi\eta$, and only the integrals over $t$ and the fermionic components of $\Psi$ are left:
\begin{align}\label{Fsimplify}
    \langle F_{11}(z)\rangle&=1+ z\,\mathcal{N}\int_{0}^{\infty} dt\exp[-c(1-e^{-t})] \int d\xi d\eta \,\exp\left[ie^{-t}g_*(\xi\eta)\right]\\
&=1-i z\,\mathcal{N}\del_\rho g_*\big|_{\rho=0,w=1/z}\int_{0}^{\infty} dt\, e^{-t}\exp[-c(1-e^{-t})]=1-i z\,\tilde{\mathcal{N}}\del_\rho g_*\big|_{\rho=0,w=1/z},\nn
\end{align}
where in the last step, we absorbed the purely numerical factor arising from the $t$-integral into the redefinition $\mathcal{N}\to\tilde{\mathcal{N}}$.
As discussed at the end of section~\ref{seqaux}, $\tilde{\mathcal{N}}$ should be fixed at the end of the computation so that $\langle F_{11}(z=0)\rangle=0$. 

\section{Asymptotic estimates}\label{seqest}

What remains to be done for completing the estimate of first return probabilities based on (\ref{Fsimplify}) is to recover some information on the saddle point solution $g_*$ from equation (\ref{sddlrho}). The complexity of this equation makes it necessary to resort to asymptotic analysis in terms of $1/c$ expansions.

To make the large $c$ structure more apparent, we employ the following rescalings:
\begin{equation}\label{rescale}
    w=\frac{\tilde w\sqrt{c}}{c+1},\qquad \rho=\frac{\tilde\rho}{\sqrt{c}},\qquad g_*=\tilde g_*-\frac{c\,\tilde w\tilde \rho}{c+1}.
\end{equation}
With these substitutions and tildes dropped for convenience, (\ref{sddlrho}) becomes
\begin{equation}
    g_*(\rho)=-ic\left(e^{-i{\rho w}/(c+1)}-1+\frac{i\,w\rho}{c+1}\right)+i\rho\, e^{-i{\rho w}/(c+1)}\!\!\int_0^\infty\! d{\rho'} \,\frac{J_{1}(2\sqrt{{\rho\rho'}/{c}})}{\sqrt{{\rho\rho'}/{c}}}e^{ig_*(\rho')-iw\rho}.\label{gsaddle1}
\end{equation}
Using $J_{1}(2x)/x\approx 1-\frac{x^2}{2}+\frac{x^4}{12}$, we can expand the right-hand side up to order $1/c^2$ as
\begin{align}\label{eqg}
&\hspace{-2mm}g_*(\rho)  
    =i\rho\int d\rho' \; e^{ig_*-iw\rho'}+\frac{\rho^2}{c}\left[\frac{ic^2w^2}{2(c+1)^2}+\frac{cw}{c+1}\int d\rho' \;e^{ig_*-iw\rho'}-\frac{i}{2}\int d\rho' \; \rho'e^{ig_*-iw\rho'}\right] \\
    &\hspace{5mm}+\frac{\rho^3}{c^2}\left[\frac{w^3}{6}-\frac{iw^2}{2}\int d\rho' \;e^{ig_*-iw\rho'}-\frac{w}{2}\int d\rho' \; \rho'e^{ig_*-iw\rho'}+\frac{i}{12}\int d\rho'\;\rho'^2 e^{ig_*-iw\rho'}\right]+\cdots, \nn
\end{align}
where we have replaced $c/(c+1)$ with 1 in the last line, as the difference would only give a contribution of order $1/c^3$. The large $c$ structure is immediately apparent from this formula: $g_*(\rho)$ can be thought of as a polynomial with $g_*(0)=0$, and the coefficient of $\rho^k$ behaves as $O(1/c^{k-1})$ at large $c$.

For computations at order $1/c^2$, we can then assume
\beq\label{guvf}
g_*(\rho,w)=-\rho\,u(w)+\frac{\rho^2}{c}v(w)+\frac{\rho^3}{c^2}f(w),
\eeq
and systematically neglect contributions smaller than $O(1/c^2)$ while solving (\ref{eqg}). Note that, at this stage, $u$, $v$ and $f$ are still unknown functions of $c$. By inserting the $c$-factors into the above formula, we have simply normalized these functions so that they are of order 1 at large $c$.
To evaluate the integrals in (\ref{eqg}), we approximate
\beq
e^{ig_*(\rho')-iw\rho'}=e^{-i\rho(u+w)}\left(1+\frac{iv}{c}\rho^2+\frac{if}{c^2}\rho^3-\frac{v^2}{2c^2}\rho^4+\cdots\right),
\eeq
and then use
\beq
\int_0^\infty d\rho\,\rho^k e^{-i\rho(u+w)}=\frac{k!}{i^{k+1}(u+w)^{k+1}}.
\eeq
With this, (\ref{eqg}) reduces to
\begin{eqnarray}
    -u&=&\frac{1}{u+w}-\frac{2iv}{c(u+w)^3}-\frac{6f}{c^2(u+w)^4} -\frac{12v^2}{c^2(u+w)^5}, \label{usol} \\
    v&=&\frac{ic^2w^2}{2(c+1)^2}-\frac{icw}{c+1}\left[\frac{1}{u+w}-\frac{2iv}{c(u+w)^3}\right]-\frac{i}{2}\left[-\frac{1}{(u+w)^2}+\frac{6iv}{c(u+w)^4}\right],
\label{vsol} \\
    f&=&\frac{w^3}{6}-\frac{w^2}{2(u+w)}+\frac{w}{2(u+w)^2}-\frac{1}{6(u+w)^3}.\label{f2}
\end{eqnarray}
At leading order, $v$ and $f$ can be neglected in (\ref{usol}), yielding 
\beq\label{Wdef}
u=W,\qquad W\equiv\frac{-w+\sqrt{w^2-4}}2,\qquad W=-\frac1{W+w}.
\eeq
There are many ways one could proceed from here with incorporating $1/c$ corrections. Most naively, we could expand $u$, $v$ and $f$ as power series in $1/c$. This approach, however, has many disadvantages. In \cite{laplace}, where the purpose was to recover the values of $g_*(\rho,w)$ at finite $w$, the singularities of $W$ at $w=\pm 2$ were seen to wreak havoc at higher orders of the naive perturbative expansion, making it necessary to expand with more care.
In our present setting, only the Laurent expansion of $u$ at $w=\infty$ is necessary to extract the first return probabilities from (\ref{Fsimplify}), so the shortcomings of the naive expansion are not so dramatic. Nonetheless, expanding blindly will produce growing numerical coefficients of higher orders in $1/c$ that would compromise numerical suppression of higher-order contributions by $1/c$. This deterioration of perturbative expansions is similar to secular terms in dynamical perturbation theory \cite{secular}. We shall therefore proceed with an approach similar to \cite{laplace}, but push our treatment to one order higher than what was done there, including contributions of order $1/c^2$.

At order $1/c$, we can neglect $f$ and $v^2$ in (\ref{usol}) and $v$ on the right-hand side of (\ref{vsol}), and furthermore replace $u$ in (\ref{vsol}), and in terms with $1/c$ in (\ref{usol}), with its lowest-order form given by $W$. This gives 
\beq
-u(u+w)=1-\frac{2iv}{c(W+w)^2},\qquad
    v=\frac{i}{2}\left(\frac{cw}{c+1}-\frac{1}{W+w}\right)^2.
\eeq
The difference between $c/(c+1)$ and 1 in $v$ could only yield corrections of order $1/c^2$ and can be neglected at order $1/c$, and then, keeping in mind that $1/(W+w)=-W$, we get
\beq\label{u1eq}
u(u+w)+1+\frac{1}{c}=0,
\eeq
consistent with the derivations in \cite{laplace}. The corresponding solution for $u$ at first order in $1/c$ is
\beq
u^{(1)}=\frac{-w+\sqrt{w^2-4(c+1)/c}}2.
\eeq
Note the `resummed' form of this expression, in the sense that we have avoided using a naive power series in $1/c$. This strategy has previously worked very well in the context of \cite{laplace}.

Finally, at order $1/c^2$, we can replace $u$ with $W$ in (\ref{f2}) and in terms of order $1/c$ in (\ref{vsol}). Then, with $1/(W+w)=-W$, 
\begin{align}\label{f2}
    f&=\frac{(W+w)^3}{6},\\
v&=\frac{i}2\left(\frac{cw}{c+1}-\frac{1}{u+w}\right)^2+\frac{2vwW^3}{c+1}+\frac{3vW^4}{c}.
\end{align}
In the last line, we can replace $u$ with $u^{(1)}$ while keeping $v$ correct at order $1/c$, and $c+1$ by $c$ in the term involving $W^3$. We then use $1/(u^{(1)}+w)=-u^{(1)}c/(c+1)$, as follows from (\ref{u1eq}), to obtain
\beq\label{v1}
v=\frac{ic^2(u^{(1)}+w)^2}{2(c+1)^2\left[1+(2W^2-W^4)/c\right]},
\eeq
since $2wW^3+3W^4=2W^3(W+w)+W^4=-2W^2+W^4$. Then, in (\ref{usol}), we can replace $u$ with $W$ in all terms of order $1/c^2$, and with $u^{(1)}$ in all terms of order $1/c$, producing
\beq
-u(u+w)=1-\frac{2iv}{c(u^{(1)}+w)^2}-\frac{6f}{c^2(W+w)^3} -\frac{12v^2}{c^2(W+w)^4}.
\eeq
We can substitute (\ref{v1}) for $v$ in the second term on the right-hand side, and $i(W+w)^2/2$ in the last term, as well as substituting (\ref{f2}) for $f$, yielding
\beq
-u(u+w)=1+\frac{c^2}{(c+1)^2(c+2W^2-W^4)}+\frac{2}{c^2}.
\eeq
The solution at order $1/c^2$ is then
\begin{align}\label{u2}
u^{(2)}&=\frac12\left(-w+\sqrt{w^2-4\left(1+\frac{c^2}{(c+1)^2(c+2W^2-W^4)}+\frac{2}{c^2}\right)}\right)\vspace{2mm}\\
&\approx\frac{-w+\sqrt{w^2-4(c+1)/c+4(2W^2-W^4)/c^2}}2,\nn
\end{align}
with $W$ given by (\ref{Wdef}).

Finally, we need to substitute (\ref{u2}) into (\ref{guvf}) and undo the rescalings (\ref{rescale}) to obtain
\beq\label{gfinal}
g_*(\rho,w)=\frac{\rho w(1-c)}{2}-\frac{\rho}{2}\sqrt{w^2(c+1)^2-4(c+1)+\frac{4}{c}[2W^2(w\sqrt{c})-W^4(w\sqrt{c})]}+\cdots,
\eeq
where we have omitted terms that vanish faster than $\rho$ at $\rho=0$, since they cannot contribute to the evaluation of (\ref{Fsimplify}).

\section{Return probability estimates}\label{seqprobs}

We proceed to substitute (\ref{gfinal}) into (\ref{Fsimplify}) to obtain
\beq
 \langle F_{11}(z)\rangle=1-i \,\tilde{\mathcal{N}}\left(\frac{(1-c)}{2}-\frac{1}{2}\sqrt{(c+1)^2-4(c+1)z^2+4z^2(2Z-cZ^2)}\right),
\eeq
where we have introduced for convenience
\beq\label{Zdef}
Z\equiv\frac{W^2(\sqrt{c}/{z})}{c}=\frac{1}{4c}\left(-\sqrt{c}/{z}+\sqrt{c/z^2-4}\right)^2=\frac{4z^2}{c^2}\left(1+\sqrt{1-4z^2/c}\right)^{\!\!-2}.
\eeq
We have $Z\sim z^2$ at $z=0$. Then, the content of the large round brackets in the above formula tends to $-c$ at $z=0$. By the discussion at the end of section~\ref{seqaux} we must impose $\tilde{\mathcal{N}}=i/c$ so that
$\langle F_{11}(z=0)\rangle=0$. As a result, we get
\beq\label{FFinal}
 \langle F_{11}(z)\rangle=\frac{c+1}{2c}\left(1-\sqrt{1-\frac{4z^2}{c+1}+\frac{4z^2(2Z-cZ^2)}{(c+1)^2}}\right),
\eeq
This formula is the final practical output of our considerations. As $Z\sim z^2$ at $z=0$, the terms involving $Z$ contribute nothing to the $z^2$ term of the Taylor expansion of (\ref{FFinal}), so that the first return probability after 2 steps is given by $1/c$ (as it would have also been on $c$-regular graphs). 

To elucidate the structure of (\ref{FFinal}), it is convenient to neglect first the subleading terms involving $Z$ and consider
\beq
 \langle F_{11}(z)\rangle_{\mbox{\tiny first-order}}=\frac{c+1}{2c}\left(1-\sqrt{1-\frac{4z^2}{c+1}}\right),
\eeq
which would have resulted from doing all the expansions to order $1/c$ rather than $1/c^2$ in section~\ref{seqest}. This relatively simple expression can be straightforwardly converted into first return probabilities by recalling the generating function of Catalan numbers $C_n$:
\beq
C(x)\equiv \sum_{n=0}^\infty C_n x^n=\frac{1-\sqrt{1-4x}}{2x},\qquad C_n\equiv\frac{1}{n+1}{2n\choose n}.
\eeq
With these relations,
\beq
 \langle F_{11}(z)\rangle_{\mbox{\tiny first-order}}=\frac{z^2}{c}\,C\!\left(\frac{z^2}{c+1}\right),
\eeq
So, return probabilities after $2n$ steps are explicitly given by
\beq \label{probsfirstorder}
\langle F_{11}(2n)\rangle_{\mbox{\tiny first-order}}=\frac{C_{n-1}}{c(c+1)^{n-1}}.
\eeq
Return probabilities after an odd number of steps manifestly vanish, which is related to \ER graphs being tree-like in the sense of having no short loops. For example, the average number of triangles adjacent to a given vertex in an \ER graph scales like $c^3/N$, evidently vanishing at large $N$.
Since returns after an odd number of steps are impossible on a tree (each edge must be traversed an even number of times), it is intuitive that our $N\to\infty$ analysis predicts vanishing return probabilities after an odd number of steps.

The first-order expression (\ref{probsfirstorder}) is strongly reminiscent of the corresponding result on $c$-regular graph derived in \cite{TBK2} that reads
\beq\label{Fcreg}
\langle F_{11}(2n)\rangle_{\mbox{\tiny $c\,$-regular}}=\frac{C_{n-1}(c-1)^{n-1}}{c^{2n-1}}.
\eeq
Note, however, that the precise way $c$ enters is different, which reflects the nonvanishing degree variance on \ER graphs. As $c$ increases, the degree variance becomes relatively smaller and smaller, and the expressions for first return probabilities on \ER and $c$-regular graphs become more and more similar.
The form of (\ref{Fcreg}) coincides with the corresponding expression on an infinite $c$-regular (Cayley) tree, which is again understood from the fact that random $c$-regular graphs are approximately tree-like. On a c-regular tree, one can choose the starting vertex as the center, and then the remaining vertices can be marked by `levels' computed as the minimal number of steps required to reach them from the central vertex. Each vertex at level 1 or higher is connected by 1 edge to the lower level and $c-1$ edges to the higher level. Hence, at each moment, the walker moves level-up with probability $(c-1)/c$ and level-down with probability $1/c$, except for the first step when it moves level-up with probability 1. To have a first return after $2n$ steps, the walk must take the first step level-up plus $n-1$ additional steps level-up, and $n$ steps level-down. Furthermore, the number of possible arrangements of these level-up and level-down steps is precisely given by the Catalan number $C_{n-1}$ since, in order for the return after $2n$ steps to be the first return, the number of steps level-up should stay strictly greater than the number of steps level-down at all times before $2n$, and counting such discrete paths is one of the standard combinatorial interpretations of the Catalan numbers. Assembling these factors together gives (\ref{Fcreg}), while many further details can be found in \cite{TBK2}. The trajectories on a tree are characterized by the property that they traverse each edge an even number of times, and the term `retroceding' has been used in \cite{TBK2} to describe this property.

Finally, the generating function of first return probability computed at order $1/c^2$ and given by (\ref{FFinal}) cannot be elegantly expanded in a closed-form Taylor series. Nonetheless, the prediction for the first return probability after any concrete number of steps can be extracted from (\ref{FFinal}) by
a brute-force Taylor expansion. The structure of this expansion is quite apparent: since $Z$ of (\ref{Zdef}) can be written as $1/c$ times a function of $z^2/c$, $4z^2(2Z-cZ^2)/(c+1)^2$ appearing in (\ref{FFinal}) is a function of $z^2/c$, divided by $(c+1)^2$. From this, it follows that contributions from the $Z$-terms to the generating function (\ref{FFinal}) will be suppressed at least as $z^{2n}/c^{n+2}$, and hence their contribution to the first return probability after $2n$ steps will be at most of order $1/c^{n+2}$, suppressed by $1/c^2$ relative to the dominant order of (\ref{probsfirstorder}), given by $1/c^n$.

In order to validate (\ref{FFinal}), we have run numerical simulations via generating a large random graph and running random walks starting from all of its vertices with a large number of trials. (We exclude vertices of degree 0, but do not otherwise track whether the vertices belong to the giant component of the graph.) This endeavor meets a few challenges: first, there are statistical uncertainties and one must collect a very large number of samples for a reliable estimate of probabilities. Second, all of our results are derived at $N\to\infty$, so we must ensure that the graphs are large enough. For example, the probability
that a vertex is adjacent to a triangle is $\sim c^3/N$. It vanishes at large $N$, and to faithfully approximate this $N\to\infty$ limit, we must have $c^3\ll N$. But that means that we should choose a large $N$ and even then a relatively small range of $c\ll \sqrt[3]{N}$ will be in a regime close to our asymptotic analysis. At the same time, $c$ cannot be too small, as that would compromise the validity of asymptotic expansions in terms of $1/c$
that has been employed in our analytics. In practice, using a sparse format for the adjacency matrix (a list of nonzero entries), we could perform our simulations at $N=1000000$, with 50000 of random walk trials per vertex. We find it useful to compare these results with our analytics by contrasting them with, first, the leading-order approximation to (\ref{probsfirstorder}) given by $C_{n-1}/c^n$, second, with (\ref{probsfirstorder}), and finally with the numbers obtained by Taylor-expanding our best approximation (\ref{FFinal}). We summarize the output of these comparisons in the following tables for $c=8$:
$$
\begin{tabular}{|l|l|l|l|l|l|}
    \hline
    & 2 steps& 4 steps& 6 steps& 8 steps &10 steps\\

        \hline
       0th order & 0.125 & 0.0156 & 0.0039 & 0.00122 & 0.000427\\
     \hline
       1st order & 0.125 & 0.0139 & 0.0031 & 0.000857&0.000267 \\
       \hline
        2nd order & 0.125 & 0.01345 & 0.0029 & 0.000787 &0.000239\\
       \hline  
       \text{numerics} & 0.12507 & 0.01335 & 0.0029 & 0.000792& 0.000243\\
    \hline
\end{tabular}
$$
and for $c=11$:
$$
\begin{tabular}{|l|l|l|l|l|l|}
    \hline
    & 2 steps& 4 steps& 6 steps& 8 steps &10 steps\\

        \hline
       0th order & 0.09091 &0.00826 &0.0015 &0.000342  &0.000087\\
     \hline
       1st order &0.09091 &0.00758  &0.00126  &0.000263 &0.0000614\\
       \hline
        2nd order &0.09091 & 0.00745 & 0.00122 & 0.000251 &0.0000577\\
       \hline  
       \text{numerics} & 0.09092& 0.00743 & 0.001224 & 0.000253&0.0000591\\
    \hline
\end{tabular}
$$
with the Catalan numbers $C_0=1$, $C_1=1$, $C_2=2$, $C_3=5$ and $C_4=14$ used for the zeroth-order and first-order estimates.
The second order, corresponding to our main result (\ref{FFinal}), evidently captures the numerics very well.

\section{Outlook}\label{concl}

We have adapted the technology of statistical field theory, in the form often used for analyzing spectral problems on random graphs,
to the question of estimating the probabilities that the first return to the original position of a random walk on an \ER graph with $N$ vertices and mean degree $c$ 
occurs after a given number of steps. Starting with the exact expression (\ref{Fjjadj}) in terms of the graph adjacency matrix,
we reduced the computation to analyzing the supervector model (\ref{SGausavC}-\ref{Cdef}). At large $N$, the solution of this model
is determined by the saddle point equation (\ref{sddlrho}). This equation is identical to the one that arises from the problem of finding
the spectral density of the normalized graph Laplacian, though the way the solution of this equation needs to be processed is different
in our present setting. Given the solution of (\ref{sddlrho}), the generating function of first return probabilities can be recovered from
(\ref{Fsimplify}), where the normalization constant is chosen to ensure that this function vanishes at the origin.

While little is known about exact solutions of the saddle point equation (\ref{sddlrho}), approximate solutions can be effectively
recovered by expanding the equation in powers of $1/c$. At order $1/c^2$, this process leads to the estimate (\ref{FFinal}) that
captures the results of numerical experiments at $c=8$ and $c=11$ very accurately. It would be interesting to make more direct contacts between
our results and the considerations of \cite{MK}, where a very different approach is formulated for analyzing first return times.

As we have displayed a strategy to process averages of the first return probability generating function (\ref{Fjjadj})
via statistical field theory computations, an arena opens up for further analysis of first return probabilities on random graphs
in those settings where statistical field theory methods have proved useful.

\section*{Acknowledgments}

We thank Ofer Biham and Eytan Katzav for very useful discussions about random walks on graphs, and particularly for drawing our attention to \cite{MK} and \cite{sites}.
 OE is supported by Thailand NSRF via PMU-B, grant number B13F670063. 
WH is supported by the NSTDA of Thailand under the JSTP scholarship, grant number SCA-CO-2565-16992-TH.


\end{document}